\documentclass[aps,prd,showpacs,floatfix,nofootinbib,twocolumn,10pt]{revtex4}

\usepackage{color,amsmath,amssymb,graphicx,latexsym,subfigure}
\usepackage{epstopdf,threeparttable,units,txfonts}
\usepackage{ulem}

\newcommand{\sigmav}{\ensuremath{\langle\sigma v\rangle}}

\newcommand{\gev}{\ensuremath{\,\mathrm{GeV}}}
\newcommand{\tev}{\ensuremath{\,\mathrm{TeV}}}
\newcommand\BR{\ensuremath{\rm BR}}
\newcommand\bsg{\ensuremath{ b\rightarrow s \gamma }}
\newcommand\brbsg{\ensuremath{\BR\left(\bsg\right)}}
\newcommand\bsm{\ensuremath{B_s\to\mu^+\mu^-}}
\newcommand\brbsm{\ensuremath{\BR\left(B_s\to\mu^+\mu^-\right)}}

\begin{document}

\title{Perspective of monochromatic gamma-ray line detection with the
High Energy cosmic-Radiation Detection (HERD) facility onboard 
China's Space Station}

\author{Xiaoyuan Huang$^1$\footnote{huangxiaoyuan@gmail.com}, 
Anna S. Lamperstorfer$^1$\footnote{anna.lamperstorfer@tum.de}, 
Yue-Lin Sming Tsai$^2$, Ming Xu$^3$, Qiang Yuan$^4$, Jin Chang$^5$, 
Yong-Wei Dong$^3$, Bing-Liang Hu$^6$, Jun-Guang L\"{u}$^7$, 
Le Wang$^6$, Bo-Bing Wu$^3$ and Shuang-Nan Zhang$^3$}

\affiliation{
$^1$Physik-Department T30d, Technische Universit\"at M\"unchen, James-Franck-Stra\ss{}e, D-85748 Garching, Germany\\
$^2$Kavli IPMU (WPI), University of Tokyo, Kashiwa, Chiba 277-8583, Japan\\
$^3$Key Laboratory of Particle Astrophysics, Institute of High
Energy Physics, Chinese Academy of Sciences, Beijing 100049, China\\
$^4$Department of Astronomy, University of Massachusetts, Amherst,
MA 01002, USA\\
$^5$Purple Mountain Observatory, Chinese Academy of Sciences, Nanjing 
210008, China\\
$^6$Xi'an Institute of Optics and Precision Mechanics, Chinese Academy 
of Sciences, Xi'an 710119, China\\
$^7$Center of Experimental Physics, Institute of High Energy Physics, 
Chinese Academy of Sciences, Beijing 100049, China
}

%\date{\today}

\begin{abstract}

HERD is the High Energy cosmic-Radiation Detection instrument proposed to
operate onboard China's space station in the 2020s. It is designed to detect
energetic cosmic ray nuclei, leptons and photons with a high energy
resolution ($\sim1\%$ for electrons and photons and $20\%$ for nuclei) 
and a large geometry factor ($>3\,{ m^2\,sr}$ for electrons and diffuse
photons and $>\unit[2]{ m^2\,sr}$ for nuclei). In this work we discuss the
capability of HERD to detect monochromatic $\gamma$-ray lines, based on 
simulations of the detector performance. It is shown that HERD will be one 
of the most sensitive instruments for  monochromatic $\gamma$-ray
searches  at energies between $\sim10$ to a few hundred GeV. Above
hundreds of GeV,  Cherenkov telescopes will be more sensitive due
to their large effective area. As a specific example, we show that
a good portion of the parameter space of a supersymmetric dark matter
model can be probed with HERD.

\end{abstract}

%95.35.+d: Dark matter
%95.85.Pw: gamma-ray
%95.85.Ry: Neutrino, muon, pion, and other elementary particles; cosmic rays
%96.50.S-: Cosmic rays
%96.50.sb: Composition, energy spectra and interactions
%98.38.Mz: Supernova remnants 
%98.70.Sa: Cosmic rays (including sources, origin, acceleration, and interactions)

\maketitle

\section{Introduction}

Cosmological observations have well established that dark matter 
(DM) constitutes $\sim25\%$ of the Universe's energy content and dominates 
the ordinary, baryonic matter \cite{Ade:2013zuv,Ade:2015xua}. The search for 
the DM particle becomes one of the most important tasks in the modern 
physics. The high energy monochromatic $\gamma$-ray emission would
be a ``smoking gun'' signature of particle DM \cite{Bergstrom:1988fp}. 
With the remarkable improvement of the sensitivity and energy resolution 
of the $\gamma$-ray detection by space and ground-based $\gamma$-ray 
instruments, great efforts have been undertaken to search for monochromatic $\gamma$-rays or sharp spectral features in  the past few years 
\cite{Abdo:2010nc,Ackermann:2012qk,Ackermann:2013uma,Ackermann:2015lka,
Bringmann:2012vr,Weniger:2012tx,Tempel:2012ey,Albert:2014hwa,
Abramowski:2013ax,Aleksic:2013xea}. There is no compelling 
evidence for the existence of line-like $\gamma$-rays yet.

Increasing the energy resolution is very crucial for the monochromatic 
$\gamma$-ray detection. The energy resolution of the current $\gamma$-ray
detectors, such as the Fermi Large Area Telescope (Fermi-LAT) in space and
the Imaging Atmospheric Cherenkov Telescope (IACT) arrays on the ground,
is of the order of $10\%-15\%$ for $O(100)$ GeV photons. Such a 
resolution is not enough to firmly identify a $\gamma$-ray line, when 
the photon statistics is not very high \cite{Bringmann:2012vr,
Weniger:2012tx}. The Alpha Magnetic Spectrometer (AMS-02) onboard the 
International Space Station has an energy resolution of $\sim2\%$ at 
$O(100)$ GeV, but the effective area of AMS-02 is too small to search for 
the weak signals \cite{Pilo:2014tpa}. The next generation of space-borne high 
energy cosmic ray (CR) and $\gamma$-ray detectors, including the CALorimetric
Electron Telescope (CALET)\footnote{http://calet.phys.lsu.edu/} \cite{2007NuPhS.166...43C}, the DArk
Matter Particle Explorer (DAMPE)\footnote{http://dpnc.unige.ch/dampe/} \cite{ChangJin:550} and
GAMMA-400\footnote{http://gamma400.lebedev.ru/indexeng.html}
\cite{Galper:2014pua} are designed to perform very high energy 
resolution ($\sim1\%-2\%$) detection of photons with large effective areas.
On the other hand, at TeV energies the ground-based Cherenkov Telescope 
Array (CTA, \cite{Consortium:2010bc}) will improve the capability of 
line-like $\gamma$-ray searches. See Refs.~\cite{Li:2012qg,
Moiseev:2013vfa,Bergstrom:2012vd,Li:2013faa,1475-7516-2015-09-048} for the 
expected performance of these future experiments.

The High Energy cosmic-Radiation Detection (HERD) facility onboard 
China's space station has been proposed recently \cite{Zhang:2014qga}. 
HERD is basically a five-side active calorimeter 
designed to perform high energy resolution and high statistics measurements 
of the CR nuclei, electrons and positrons, and $\gamma$-ray photons in space. 
The scientific objectives of HERD include the high sensitivity search 
for particle DM, direct measurements of the CR nuclei spectra and 
composition up to knee energies, and high energy 
$\gamma$-ray sky surveys. Based on the detailed simulations of the HERD 
detector \cite{Xu:2014raa}, we investigate the expected performance of 
HERD on the monochromatic $\gamma$-ray line detection and the potential
on the constraints of DM model parameters in this work. 

This paper is organized as follows. In Sec.~II we briefly introduce
the design and performance of the HERD detector. The sensitivity of line
searches is presented in Sec.~III. Taking the Minimal Supersymmetric 
Standard Model (MSSM) as an example, we show the capability of HERD 
to explore the corresponding DM parameter space in Sec.~IV. Finally we
conclude in Sec.~V.

\section{HERD design and performance}

\begin{figure}[!htb]
\centering
\includegraphics[width=0.48\textwidth]{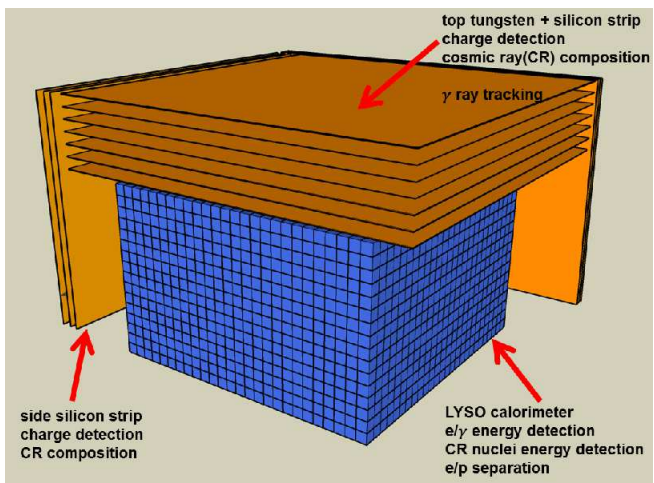}
\caption{Schematic design plot of HERD.
\label{fig:design}}
\end{figure}

HERD is composed of a 3-D cubic calorimeter (CALO) surrounded by microstrip 
silicon trackers (STKs) from five sides, while the bottom is left for 
mechanical support. Fig.~\ref{fig:design} shows a schematic plot of the 
basic design and the major functions of each part of the detector. The CALO 
is made of $21\times21\times21$ cubic LYSO crystals with each cell 
$3\times3\times3$ cm$^{3}$ which is coupled with the wavelength shifter 
fibers and read out by ICCD. This design corresponds to about 55 radiation 
lengths and 3 nuclear interaction lengths, respectively. The CALO detects 
the total energy deposited by the electromagnetic (EM) shower from CR 
leptons/photons and the hadronic shower from CR nuclei, and provides 
lepton/hadron separation through the differences of the shower shape. 
The five sides of identical STKs, each consists of seven layers laid
perpendicularly one by one, are sandwiched with tungsten converters 
(2 radiation lengths), to ensure the maximum field of 
view (FOV) and provide charge identification, trajectory measurements, 
back scattering rejection, as well as some early shower information of 
photons and electrons. The  detector is surrounded by plastic 
scintillators from  five sides, which are used to reject the low 
energy charged particles to maximize the efficiency for photons and high 
energy CRs including electrons.

Extensive simulations have been carried out with GEANT4
\cite{Agostinelli:2002hh} and FLUKA \cite{Battistoni:2007zzb} to get the scientific
performance of HERD \cite{Xu:2014raa}. We generate events, including photons,
electrons and protons, at energy grids of 0.5, 1, 5, 10, 50, 100, 200, 300,
500, 700, and 1000 GeV, with an isotropic event generator. The shower
development in the detector of each event is simulated using the Quark-Gluon
String Precompound (QGSP) interaction model for photons/electrons and
DPMJET-III model for protons. The size, orientation, and shape of the shower
is used to reconstruct the energy, direction, and type of the incident
particle. To ensure the best determination of the energies for photons and
electrons, we select those events whose shower maximum is fully sampled in
CALO, which results in a decreasing fiducial area with increasing energy.
The lepton-hadron discrimination is primarily based on the shower shape
in CALO. The charge measurement by STKs can be used to identify neutral
particles from charged ones. A machine learning, multivariate analysis
with boosted decision trees is adopted to identify the particle type.
The background (electrons and protons) rejection efficiency is obtained
when keeping 90\% signal (photons). We assume infinite shield of the
bottom side of the detector from the space station. The shield of the
other five sides is negligible. This assumption will effectively exclude
the earth limb's photons due to the fact that the space station will
point away from the earth surface for almost all of the time. The secondary
events due to interactions between incident CRs and the space station,
which would travel upwards in the detector, will also be excluded.

The ideal energy resolution of the EM shower is about 0.1\% at 200 GeV. 
Considering the stochastic fluctuation of the number of photoelectrons 
and the error of energy calibration, the energy resolution for the EM 
component can be $\sim1\%$. Due to the large nuclear interaction length 
of the CALO, the energy resolution of the hadronic component is around 
20\%, almost constant from hundreds of GeV up to PeV. The CALO can be 
also used for the lepton/hadron separation due to the fact that the EM 
and hadronic showers differ in their spatial and energy distributions 
in the high granulated crystals. The hadron efficiency is about 
$5\times10^{-6}$ when keeping $90\%$ of the EM events. The key performance 
of HERD, in comparison with previous and other proposed missions, is its 
extremely large effective geometry factor for all types of high energy 
particles, primarily due to its thick 3-D CALO and five-sided STKs. 
With homogeneous design for detecting particles from every unblocked 
direction, the effective geometry factor is $>3$ m$^{2}$sr for electron 
and diffuse $\gamma$-rays, and $>2$ m$^{2}$sr for CR nuclei.

The energy resolution of CALO can be parameterized as the quadratic addition
of the statistical fluctuation term ($\propto E^{-1/2}$) and the electronic
noise term ($\propto E^{-1}$)
\begin{equation}
\label{sigma}
\frac{\sigma_E}{E}=\left(\frac{49\,{\rm GeV}}{E}+\frac{12\,{\rm GeV}^2}
{E^2}\right)^{1/2}\%.
\end{equation}
The minimum energy resolution is taken to be $1\%$, due to the systematic
effect of the calibration and the non-uniformity. The effective exposure
depends on the orbit of the space station. Employing the orbit of Tiangong-1
in 2012-2013, we simulate the all-sky exposure map of HERD, taking the 
shield of the space station into account. The Galactic center region is 
of particular interest for the DM searches. An average exposure of $\sim 
0.65$ yr is achieved for this direction and one year of operation.
The photon detection efficiency is the product of the
$e^{\pm}$ pair conversion efficiency ($\sim80\%$), the reconstruction
efficiency ($\sim90\%$), the  efficiency for electron rejection ($90\%$ for
$10^{-3}$ residual electrons), and the efficiency for hadron rejection
($90\%$ for $10^{-8}$ residual hadrons), which is in total $\sim60\%$.
The geometry area of HERD/CALO is $63\times63$ cm$^2$ on each side. If the
energy of the incident $\gamma$-ray photon is high enough, there will
be leakage of the shower for the events hitting the edge of the detector.
The fiducial area for good events can be approximated as $(3660-717\times
\log(E/{\rm GeV}))$ cm$^{2}$, for the energies from $\sim10$ GeV to 
$\sim$TeV. Considering the correction of the shower leakage, the effective 
area can be larger. Here we adopt this result as a somehow conservative
estimate.

\section{Sensitivity of the line search}

\subsection{Backgrounds}

The backgrounds for monochromatic $\gamma$-ray searches include misidentified
charged particles (nuclei and leptons) and the continuous 
$\gamma$-ray events. The misclassified CRs are isotropic, due to the
loss of the direction information during the diffusive propagation.
For the proton flux, we adopt a fitting formula based on the recent 
AMS-02 data \cite{Aguilar:2015ooa}
\begin{eqnarray}
\phi_{p}(E_k)&=&7.8\times10^{-2}(E_k/{\rm GeV})^{-0.9}\nonumber\\
&\times&\left[1.0+(E_k/4.9\,{\rm GeV})^{1.87}\right]^{-1}
{\rm GeV^{-1}cm^{-2}s^{-1}sr^{-1}}.\nonumber\\
\end{eqnarray}
The nucleon flux from Helium, is lower by a factor of 
$\sim3-5$ in the energy range of $10-1000$ GeV/nucleon \cite{Abe:2015mga}.
Heavier nuclei play an even less important role compared to protons.
Since the hadron rejection power of HERD is very high, the nucleon 
background is always subdominant in the current study. The total  $e^-+e^+$ spectra observed by AMS-02 
\cite{Aguilar:2014fea} and HESS \cite{Aharonian:2008aa} can be fitted by
\begin{eqnarray}
\phi_{e^{\pm}}(E_k)&=&1.85\times10^{-3}(E_k/{\rm GeV})^{-0.71}\nonumber\\
&\times&\left[1.0+(E_k/3.5\,{\rm GeV})^{2.63-0.05\log_{10}
(E_k/{\rm GeV})}\right]^{-1}\nonumber\\
&\times&\left[1+(E_k/1300\,{\rm GeV})^5
\right]^{-0.3}{\rm GeV^{-1}cm^{-2}s^{-1}sr^{-1}}.\nonumber\\
\end{eqnarray}
The efficiencies of protons as well as $e^+$ and $e^-$, $\eta_p=10^{-8}$ 
and $\eta_e=10^{-3}$, are multiplied with the CR fluxes in order to get 
the backgrounds of the misclassified CRs. With such a
high rejection power of the CR hadrons, the hadronic background can be
safely neglected for HERD. This is different from the IACTs, for which
the hadronic background is comparable to other backgrounds
\cite{Bringmann:2011ye}.

The $\gamma$-ray sky is composed of diffuse emission and point or extended
sources. Since the bright point-like sources can be effectively removed
in the data analysis, and the residual weak sources are expected to
contribute $\lesssim10\%$ to the diffuse events \cite{Ackermann:2013uma}, 
we disregard the point sources in this analysis. Only the Galactic
and extragalatic diffuse $\gamma$-rays are taken into account as the
continuous background. 

The Galactic diffuse $\gamma$-ray emission can be modeled with three 
components that can be distinguished according to their radiation 
mechanisms: (a) the decay of
neutral pions produced by inelastic collisions of CR protons with the 
interstellar gas, (b) inverse Compton scattering of the interstellar 
soft photons by CR electrons and positrons, and (c) bremsstrahlung 
radiation produced by the scattering of CR electrons and positrons with 
the interstellar gas. There are also a few identified large scale 
structures such as the Fermi bubbles \cite{Dobler:2009xz,Su:2010qj}, 
Loop I \cite{Casandjian:2009wq}, and the Magellanic stream. 
The majority of the Fermi diffuse $\gamma$-ray emission can be relatively 
well modelled with the CR interactions, the large scale structures, the residual CR events and point sources, except for some excesses in 
the Galactic plane \cite{FermiLAT:2012aa}. Based on the physical modeling
of the Galactic diffuse $\gamma$-ray emission and a likelihood fitting
of the all-sky data, the Fermi Collaboration built diffuse $\gamma$-ray
templates for the point source analyses. Those templates are suitable to 
simulate the $\gamma$-ray background in order to forecast the sensitivity 
of line searches with future experiments. Since the Fermi data are 
limited to a few hundred GeV, we need to extrapolate the templates
to higher energies. We focus on the inner Galaxy region, where
the uncertainties of the spectra of the Fermi bubbles will affect the
extrapolations \cite{Fermi-LAT:2014sfa}. Thus we adopt two different 
templates, to take such uncertainties into account. The first template is
the one used for the ``p6v11'' data analysis{\footnote{http://fermi.gsfc.nasa.gov/ssc/data/analysis/software/aux/gll$\_$iem$\_$v02$\_$P6$\_$V11 $\_$DIFFUSE.fit}}
without the Fermi bubbles. The second one is built with the Fermi bubbles 
and is used for the Pass 7 Reprocessed Data{\footnote{http://fermi.gsfc.nasa.gov/ssc/data/analysis/software/aux/gll$\_$iem$\_$v05$\_$rev1.fit}}. 
These templates are transformed into HEALPix{\footnote {http://healpix.jpl.nasa.gov}} projections with $N_{\rm side}=64$. Then the spectrum of each pixel
is extrapolated to 2 TeV. Due to the suppression of the Klein-Nishina 
cross section, the simple extrapolation will tend to over-estimate the
inverse Compton scattering emission at high energies, which makes the
line limits more conservative.

For the extragalactic diffuse emission, we adopt the fitting formula of
the latest Fermi data, which extends up to 820 GeV \cite{Ackermann:2014usa}
\begin{eqnarray}
\phi_{\rm EG}=I_{0.1}\left(\frac{E}{0.1\,{\rm GeV}}\right)^{-\gamma}
{\rm exp} \left(-\frac{E}{E_{\rm cut}}\right)\,.
\end{eqnarray}
The parameters depend on the assumptions of the foreground model 
\cite{Ackermann:2014usa}. In this work we adopt $I_{0.1}=0.95\times10^{-4}$ 
GeV$^{-1}$cm$^{-2}$s$^{-1}$sr$^{-1}$, $\gamma=2.32$ and $E_{\rm cut}=279$
GeV, corresponding to model A in Ref.~\cite{Ackermann:2014usa}. Different choices of the parameters
will not affect our results significantly because the extragalactic background is 
only a sub-dominant contribution to the total background.

\begin{figure*}[!htb]
\centering
\includegraphics[width=0.48\textwidth]{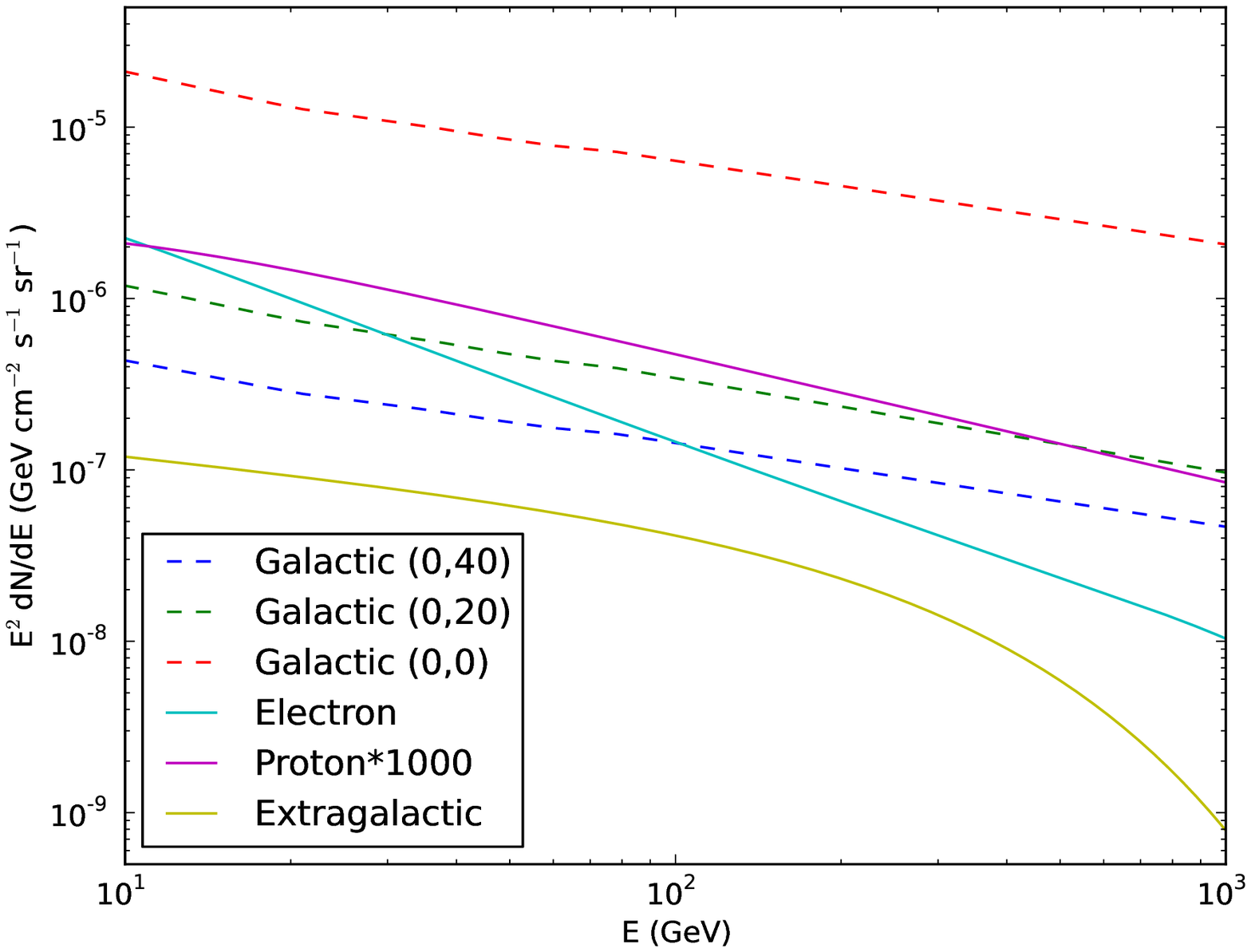}
\includegraphics[width=0.48\textwidth]{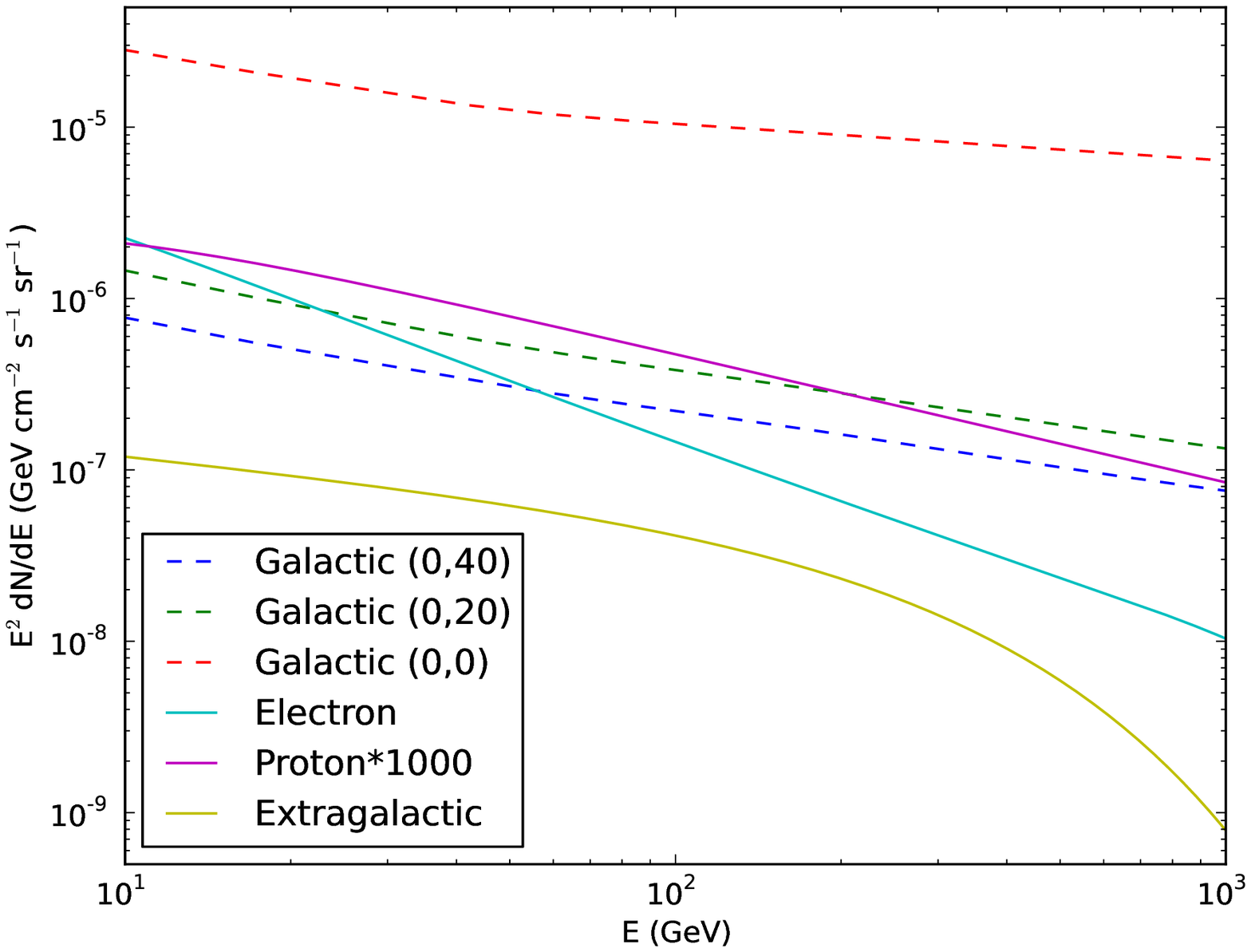}
\caption{Comparison of the different backgrounds: the Galactic and isotropic
diffuse continuous $\gamma$-ray backgrounds, CR electrons, and protons
(multiplied by $10^3$). The left panel is for ``p6v11'' template, and the
right panel is for ``p7v6'' template. The Galactic diffuse $\gamma$-ray 
emission is anisotropic, and we show the fluxes in three directions with 
$(l,b)=(0,0)$, $(0,20^\circ)$, and $(0,40^\circ)$, respectively.
\label{fig:bkg}}
\end{figure*}

Fig.~\ref{fig:bkg} shows the contributions of different components to
the background. Since the Galactic diffuse emission is anisotropic, the 
spectra are given for different directions in the sky, concretely for 
the Galactic Center, as well as for low and high latitude regions. 
We can see that at low latitudes, the contribution from the Galactic 
diffuse component dominates over the other components. When approaching 
the high latitude regions, the misclassified $e^+$ and $e^-$ become 
more and more important, especially at low energies. The fluxes of 
misclassified protons and the extragalactic $\gamma$-ray background 
are always lower than the $e^-$ and $e^+$ backgrounds.

\subsection{DM profile and region of interest}

The density profile of the DM distribution in the Milky Way halo 
comprises large uncertainties in the innermost region. As generally
adopted in the literature, we consider several typical DM density profiles
 to take such uncertainties into account. The first one is the
commonly adopted cuspy Navarro-Frenk-White (NFW) profile 
\cite{Navarro:1996gj},
\begin{equation}
  \rho(r) = \frac{\rho_s}{(r/r_s)(1+r/r_s)^2}
\end{equation}
with $r_s=20$~kpc. The second one is the Einasto profile
\cite{1965TrAlm...5...87E} with an asymptotic flat slope in the center
\begin{equation}
  \rho(r) = \rho_s \exp\{ -(2/\alpha)[(r/r_s)^\alpha - 1]\},
\end{equation}
where $r_s=20$ kpc and $\alpha=0.17$, which is favored by more recent 
simulations \cite{Navarro:2008kc}. The third one is the cored isothermal 
profile \cite{Bahcall:1980fb}
\begin{equation}
  \rho(r) = \frac{\rho_s}{1+(r/r_s)^2}
\end{equation}
with $r_s=5$~kpc. All the profiles are normalized to local
density of $\rho_\odot=0.4$ GeV cm$^{-3}$.

\begin{figure}[!htb]
\centering
\includegraphics[width=0.48\textwidth]{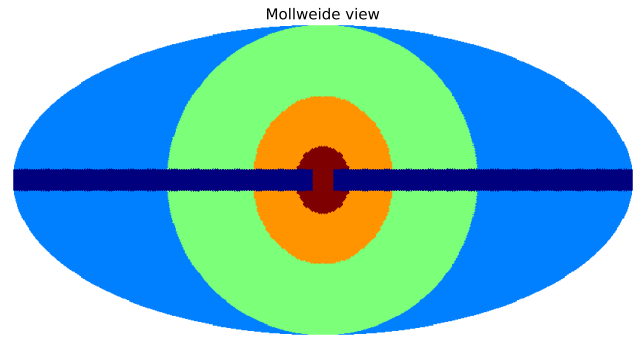}
\caption{Illustration of the different ROIs for different assumptions of the
DM density distribution. From inside to outside, the regions are R16,
R41, R90, and R180, respectively.
\label{fig:mask_line}}
\end{figure}

Optimized regions of interest (ROI) are very important  in the
search for the weak DM signal. Since the expected signal depends on
the DM density profile, we adopt different ROIs for different
assumed profiles. Following Ref.~\cite{Ackermann:2013uma}, we choose a
circular region with radius of $16^\circ$ (R16) for the Einasto profile,
$41^\circ$ (R41) for the NFW profile, and $90^\circ$ (R90) for the isothermal
profile, in the case of DM annihilation. We also discuss the 
decaying DM scenario, for which the optimized search region (R180) 
corresponds to the whole sky for all these profiles. In fact, the 
integrals of the DM density over the Milky Way are very similar among
these profiles (see Table \ref{tab:roi_jval}). For definiteness, we adopt the 
NFW profile for decaying DM. In all these ROIs the Galactic plane with 
$|l|>6^\circ$, $|b|<5^\circ$ is removed. Fig.~\ref{fig:mask_line} shows 
the ROIs adopted to search for monochromatic $\gamma$-rays in this work. 
Table \ref{tab:roi_jval} summarizes the optimized ROI for each DM halo 
profile and the corresponding $J$-factors (integral of $\rho^2$ or $\rho$ 
over the line-of-sight and ROI). The results differ slightly from that
of Ref.~\cite{Ackermann:2013uma}, due to the mask of point sources in
Fermi's paper. 
 
%\begin{table}[ht]
%\caption{\label{tab:roi_jval}Summary of the optimized ROIs and $J$-factors.}
%\begin{center}
%\begin{tabular}{ccccc} 
%\hline\hline
%Profile& ROI& $J_{\rm anni}$ & $J_{\rm dec}$ \\
% & & (10$^{22}$ GeV$^{2}$ cm$^{-5}$) & (10$^{23}$ GeV cm$^{-2}$)\\
%\hline
%Einasto         & R16  &   9.39  & \\
%NFW             & R41  &   9.17  & \\
%Isothermal      & R90  &   6.95  & \\
%\hline
%Einasto            & R180 &         & 2.55\\
%NFW             & R180 &         & 2.52\\
%Isothermal            & R180 &         & 2.56\\
%\hline\hline
%\end{tabular}
%\end{center}
%\end{table}

\begin{table}[htb]
\caption{\label{tab:roi_jval}Summary of the optimized ROIs and $J$-factors.}
\begin{center} 
\begin{tabular}{  l  c  c c c } 
\hline 
\hline
Profile & ROI & $J_\text{ann}$ &  ROI & $J_\text{dec}$ \\ 
& & ($\unit[10^{22}]{GeV^2 \, sr \, cm^{-5}}$)&& ($\unit[10^{23}]{GeV \,sr \,cm^{-2}}$)\\
\hline 
Einasto    & R16 & 9.39 & R180 & 2.55  \\ 

NFW        & R41 & 9.17 & R180 & 2.52  \\
Isothermal & R90 & 6.95 & R180 & 2.56   \\ 
\hline 
\hline
\end{tabular} 
\label{J-factors:Herd}
\end{center}
\end{table}

\subsection{Mock Data and statistical treatment}

Given the expected $\gamma$-ray fluxes including the CR backgrounds and
the performance of the instrument, we can generate mock data for HERD. 
We set 400 energy bins per decade in the energy range $\unit[5]{GeV}-
\unit[2]{TeV}$, and calculate the expected number of counts 
$n_{\rm exp}^{i}$ in each energy bin $i$ with width $\Delta E_{i}$ is
\begin{equation}
\label{mockdata}
n_{\rm exp}^{i}=\Delta t \int_{\Delta E_{i}}dE
\int dE'R(E,E') A_{\rm eff}(E')\Phi_{\rm tot}^{\rm roi}(E'),
\end{equation}
where $\Delta t$ is the exposure time, $R(E,E')$ is the energy response 
function of the instrument, $A_{\rm eff}(E')$ is the effective area, 
and $\Phi^{\rm roi}_{\rm tot}$ is the total $\gamma$-ray and CR background
in the ROI. In this work we adopt 5  years of survey time, which 
corresponds to an average $\sim3.25$ year effective exposure. The energy 
response function is assumed to be Gaussian, and its width is given in 
Eq.~(\ref{sigma}). Assuming background only, we generate the mock 
observational counts $n_{\rm obs}^{i}$ in each energy bin, by generating 
random numbers that are drawn from a Poisson distribution with expectation
$n^i_\text{exp}$. Here we neglect the effect of the point spread function 
(PSF), which is expected to be small since we integrate the  $\gamma$-ray fluxes 
in large enough sky regions compared to the resolution angle.

Then we calculate the expected number of counts of the theoretical model
with the DM contribution. The monochromatic $\gamma$-ray flux from DM 
annihilations or decays reads
\begin{equation}
\Phi^{\rm roi}_{\rm DM}(E)=N_{\gamma}\left\{
\begin{array}{ll}
\frac{\sigmav\times J_{\rm anni}}{8\pi m_{\chi}^2}\times R(m_\chi,E), & \textrm{annihilation}\\
\frac{J_{\rm dec}}{4\pi m_\chi \tau}\times R(m_\chi,E), & \textrm{decay}
\end{array}\right.
\end{equation}
in which $\sigmav$ or $\tau$ is the annihilation cross section or decay
lifetime of the DM particle, $m_\chi$ is the DM mass, and $N_{\gamma}$
is the multiplicity of one annihilation or decay. On the other hand, 
the theoretical background is parameterized as a single power-law 
function $\Phi_{\rm bkg}^{\rm roi}=CE^{-\gamma}$, which is good enough 
to approximate the background spectrum in a narrow energy window (see 
below the definition of the energy window). Our model for the $\gamma$-ray 
flux contains thus three parameters, the two background parameters $C$ 
and $\gamma$, as well as the flux from DM annihilation or decay (proportional
to $\sigmav$ or $1/\tau$). The expected number $n^i_\text{exp}$ of counts 
are calculated by substituting the background with our three parameter 
flux model, $\Phi_{\rm bkg}^{\rm roi}+\Phi^{\rm roi}_{\rm DM}$, in 
Eq.~\eqref{mockdata}. The likelihood function is given by
\begin{equation}
\label{likelihood}
\mathcal{L}\propto \prod_i \frac{\left(n^i_\text{exp}\right)^{n^i_\text{obs}} \text{exp}(-n^i_\text{exp}) }{n^i_\text{obs}!}.
\end{equation}

We adopt the profile likelihood analysis \cite{Rolke:2004mj} and implement
the sliding window technique \cite{Weniger:2012tx}, in order to derive the 
upper limits of the DM flux. 
We follow Ref.~\cite{1475-7516-2015-09-048} to define the energy 
windows. For a specified central energy $\bar{E}$, which is the energy of 
the $\gamma$-ray line to be searched, the energy window is defined to be 
$[\bar{E}/\sqrt{\epsilon},[\bar{E}\sqrt{\epsilon}]$. The parameter
$\epsilon$ is determined to ensure that the astrophysical background is 
well described by a power law. Specifically, we employ $\chi^2$ fittings 
to the 300 mock data sets with a power-law model, for each ROIs and for
different window sizes (from $\epsilon=1.2$ to $\epsilon=8$). The resulting
distribution of the 300 $\chi^2$ values for each ROIs is then tested against a $\chi^2$
distribution with the number of degrees of freedom given by the number
of energy bins in the energy window minus 2 (number of fitting parameters).
An energy window is rejected if the corresponding p-value is smaller than 0.01. And our chosen energy window corresponds to the largest window that fulfills that criterion for all ROIs under consideration. This procedure gives $\epsilon\sim1.6$ at
20 GeV and $\sim4$ at 1 TeV. We have tested that adopting the parameter
$\epsilon$ twice as large as the above derived values, the limits improve
by only $7\%$ above 40 GeV. A possible optimization is to choose different
$\epsilon$ for different ROIs. However, the results will not change
significantly.

Maximizing the logarithm of Eq.~(\ref{likelihood}) with respect to the 
two background free parameters ($C$, $\gamma$) and one DM parameter 
$\Phi_{\rm DM}^{\rm roi}$, we find the best-fit log-likelihood 
$\log\mathcal{L}_{\rm bf}$. The one-sided 95\% confidence level (CL) 
upper limit on the DM flux is found by increasing the signal starting 
from the best fit value until $-2\log \mathcal{L}(\Phi_{\rm DM}^{\rm roi})
=-2\log\mathcal{L}_\text{bf}+2.71$ \cite{Cowan}.

\subsection{Results}

\begin{figure}[!htb]
\centering
\includegraphics[width=0.48\textwidth]{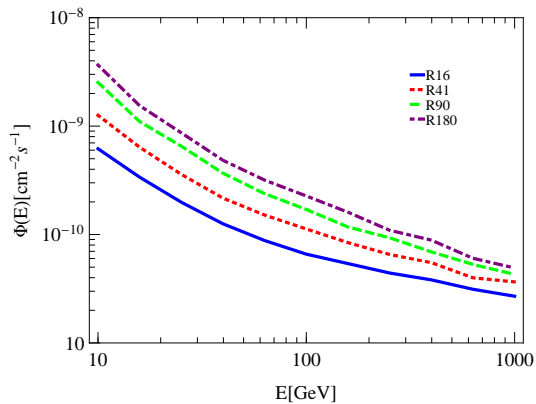}
\caption{95\% upper limits on the monochromatic photon fluxes for different
ROIs, where the Galactic background template ``p7v6'' is adopted.
\label{fig:FluxLimits}}
\end{figure}

We perform a scan of the photon energies from $10$~GeV to $1$~TeV, 
and calculate the profile likelihood with respect to the monochromatic
line flux for each photon energy. The $95\%$ CL upper limits of the 
line fluxes for different ROIs are shown in Fig.~\ref{fig:FluxLimits}. 
Here we adopt the ``p7v6'' Galactic background model. The results are 
the geometric mean of the limits obtained from 300 mock data realizations. 
It is shown that at low energies the differences among various ROIs are 
larger than at high energies. This reflects the effect of the relative 
weights between the electron background and the Galactic diffuse emission.
At low energies the electron background plays an important role in the
total background. Thus the backgrounds in different ROIs scale with
the solid angles of the sky regions, and hence differ significantly. At high
energies, however, the anisotropic Galactic diffuse background, which 
is dominated by low latitude emissions, becomes more and more
important, and the differences among these ROIs become smaller.

\begin{figure*}[!htb]
\centering
\includegraphics[width=0.48\textwidth]{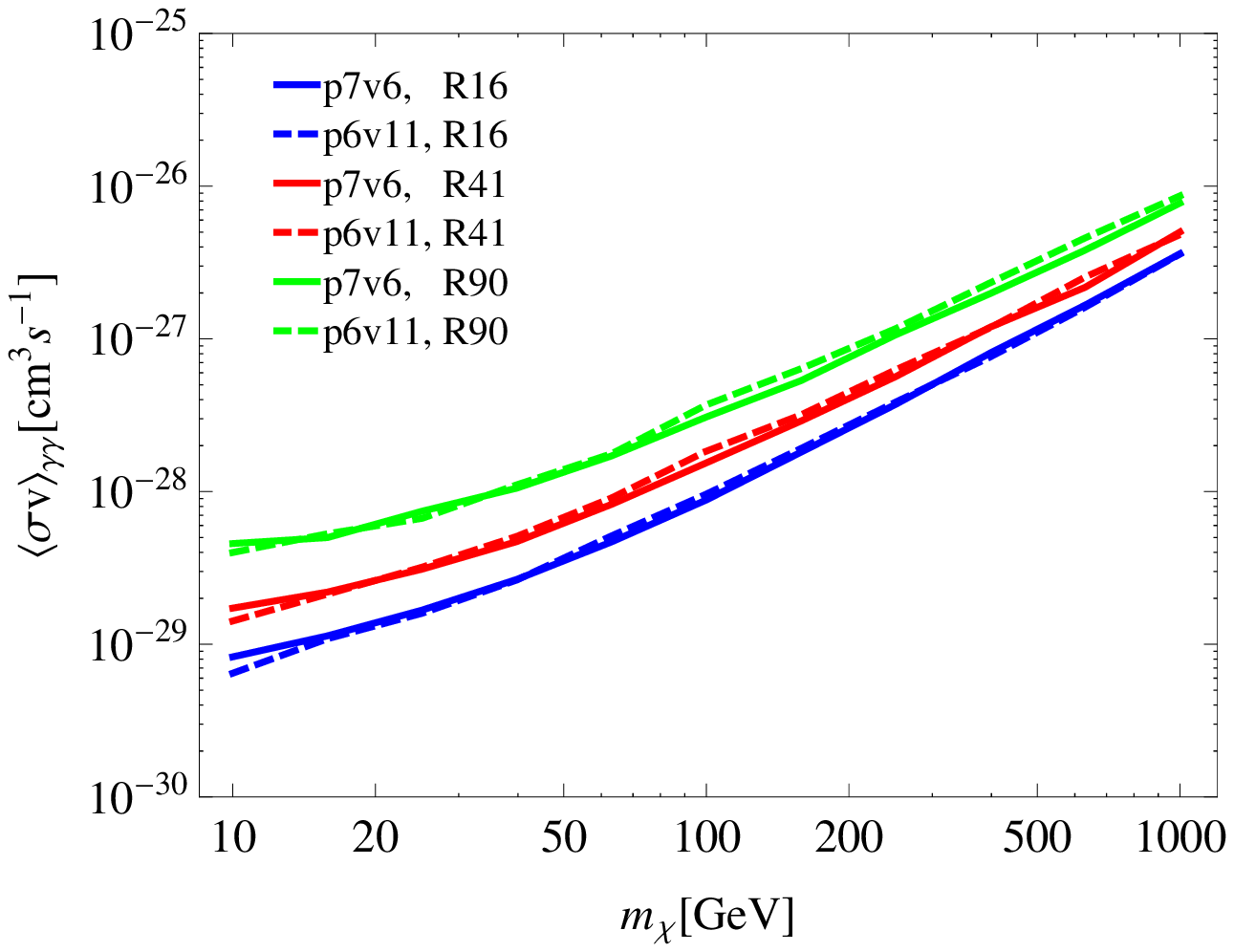}
\includegraphics[width=0.48\textwidth]{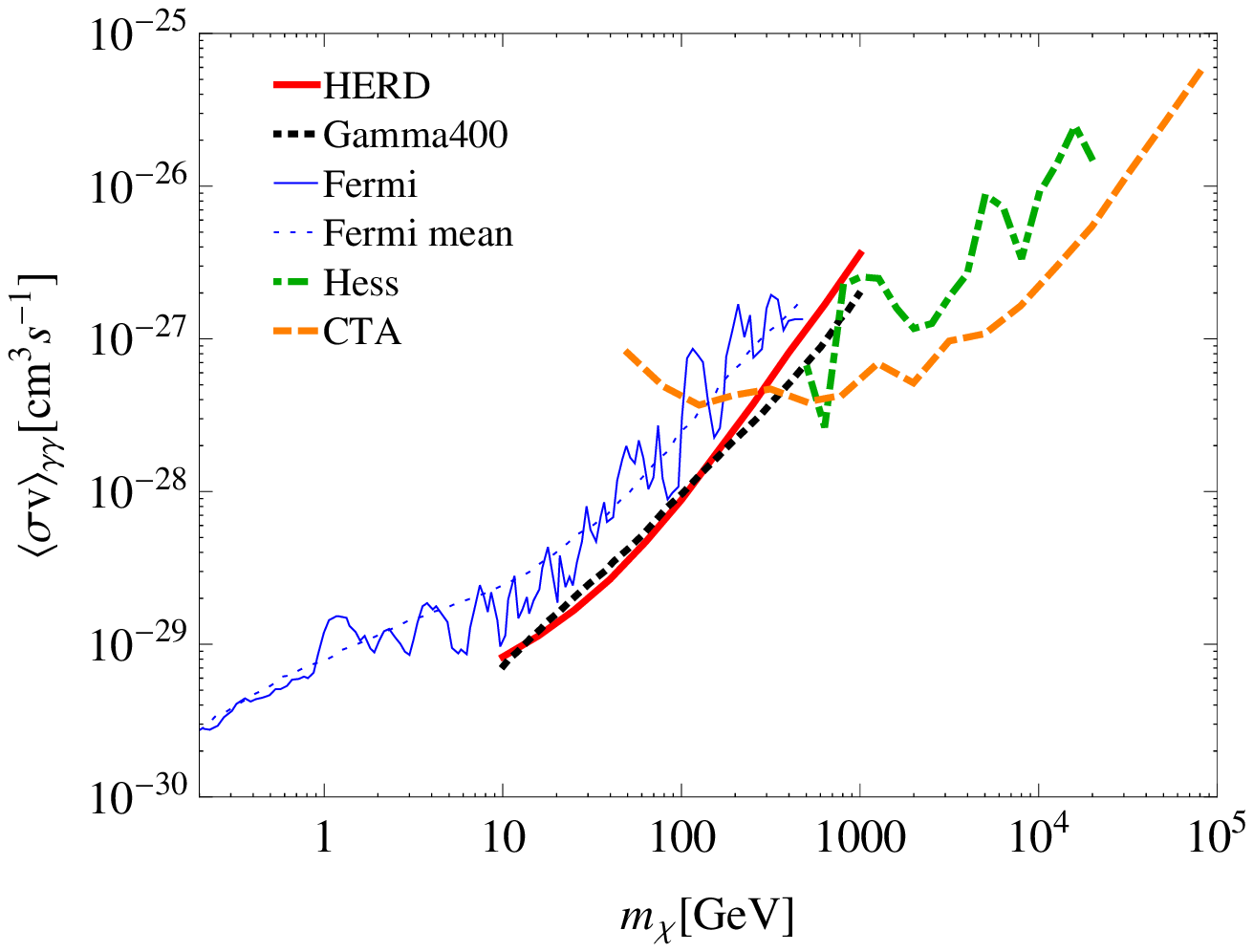}
\caption{Left panel: the mean limits obtained from 300 realizations
of mock data, for the three ROIs (DM profiles): R16 (Einasto), R41 (NFW) 
and R90 (isothermal). The solid (dashed) lines correspond to ``p7v6'' 
(``p6v11'') Galactic background. Right panel: comparison of HERD results
with limits from Fermi \cite{Ackermann:2015lka} (solid is the observational
limits and dashed is the expected mean) and HESS \cite{Abramowski:2013ax}, 
and that predicted for GAMMA-400 \cite{Bergstrom:2012vd} and CTA 
\cite{1475-7516-2015-09-048}. See the text for details.}
\label{fig:results}
\end{figure*}

The upper limits of the line fluxes can be easily translated into the
upper limits of the DM annihilation cross section or decay lifetime.
The left panel of Fig.~\ref{fig:results} shows the 95\% upper limits
on the cross section of DM annihilation into a pair of photons
$\sigmav_{\gamma\gamma}$. We find very stringent limits, reaching 
cross sections of  $6 \cdot \unit[10^{-30}] \sim 5 \cdot 
\unit[10^{-29}]{cm^3s^{-1}}$ at $10$~GeV, and  $4 \cdot 10^{-27}\sim 9 \cdot 10^{-27}$
cm$^3$s$^{-1}$ at TeV. For the two Galactic background templates we find 
very similar results in all three regions. The largest difference amounts 
to about $30\%$, due to the contribution of the Fermi bubbles to the 
background (see Fig.~\ref{fig:bkg}). We expect that below $\sim100$ GeV 
the ``p7v6'' background should describe the actual background better, while 
above $\sim100$ GeV where the energy spectra of the Fermi bubbles show a
cutoff \cite{Fermi-LAT:2014sfa}, the extrapolation of the ``p7v6'' 
template will over-estimate the background. The limits for R16 and R90 
differ by a factor of about $5$ at $10$~GeV and $2$ at $1$~TeV, even though
the $J$-factors of these regions are very similar. Such differences
come from the different background levels of these ROIs. In analogy to the
flux limits, larger differences at low energies arise due to the electron
background.

In the right panel of Fig.~\ref{fig:results} we compare the HERD limits
with the results from other current or upcoming $\gamma$-ray facilities. 
All the limits shown are for the Einasto profile. For the HERD limits
we adopt the ``p7v6'' background model. The Fermi
results are adopted from Ref.~\cite{Ackermann:2015lka}, with the same
ROI (R16) and the newest analysis of the Pass 8 data. The prediction of 
GAMMA-400 is from a $20^\circ$ region around the Galactic center, 
excluding the Galactic disc ($|l|>5^\circ$ and $|b|<5^\circ$), for 5 years 
of full sky survey \cite{Bergstrom:2012vd}. It is shown that HERD can 
improve the Fermi limits \cite{Ackermann:2015lka} from 5.8 years of 
observation by up to a factor of a few. The limits expected from 
GAMMA-400 and HERD are comparable. However, the effective area of 
GAMMA-400 used in Ref.~\cite{Bergstrom:2012vd}, as well as the 
$\gamma$-ray efficiency, seems to be too ideal \cite{Galper:2014pua}.
The IACTs are expected to be more effective to probe the $\gamma$-ray lines 
at high energies. We also show the limits from 112 h of Galactic center 
observation with the HESS telescopes \cite{Abramowski:2013ax} in a circular 
$1^\circ$ region around the GC, excluding $|b|<0.3^\circ$, as well as the 
expected limits for CTA \cite{1475-7516-2015-09-048}. The CTA limits are derived 
for the same region and observation time as that of HESS \cite{1475-7516-2015-09-048}
and are properly rescaled to $\gamma$-ray lines.
We can see that below $\sim300$ GeV HERD is more sensitive than CTA, whereas
at higher energies CTA is expected to be more sensitive.

\begin{figure}[!htb]
\centering
\includegraphics[width=0.48\textwidth]{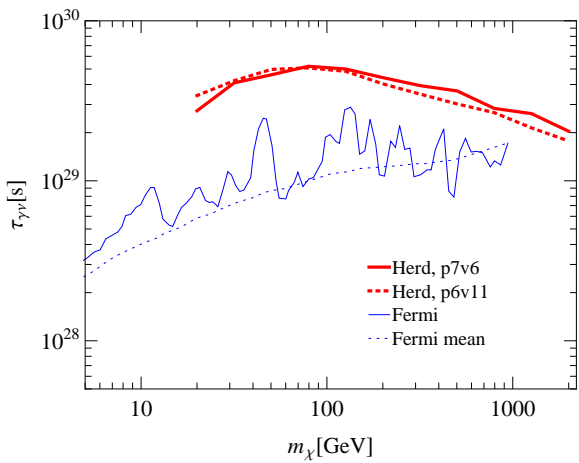}
\caption{95\% upper limits of the decay lifetime $\tau_{\gamma\nu}$ of the 
DM particle. The observed and expected mean limits from 5.8 years of Fermi 
observations are also shown for comparison \cite{Ackermann:2015lka}. 
\label{fig:DecR180NFW}}
\end{figure}

The constraints on the lifetime of DM decay into $\gamma\nu$ are shown
in Fig.~\ref{fig:DecR180NFW}. Here we adopt the NFW profile and the 
R180 region. For the sake of comparison with Fermi, we show the limits 
on $\tau_{\gamma \nu}$, for $\chi\to\gamma\nu$ channel. A classical
example of this kind of DM is the gravitino in the supersymmetric
model \cite{Takayama:2000uz}. Similarly, we find that the HERD limits 
can improve the current Fermi-LAT constraints \cite{Ackermann:2015lka} 
by a factor of $\sim 2-3$. 

\section{Constraints on MSSM parameters}

In this section we show the HERD potential on the constraints of the
specific MSSM DM model parameter space. Since we are only interested in 
the $\gamma$-ray line features, we focus on the lightest neutralino in 
the MSSM, the DM candidate, that can monochromatically annihilate to 
$\gamma\gamma$ and $\gamma Z$ via loop processes. Because of the 
loop suppression, the cross section is usually rather small compared to 
the total annihilation cross section. On the other hand, we want to 
demonstrate the power of HERD in the higher cross section region. 
Therefore we ignore the sfermion-DM coannihilation 
region that exhibits low annihilation cross sections. The mass splitting 
between a sfermion and DM is small enough to release the Boltzmann 
suppression so that the correct relic density can be reached by 
accounting for neutralino-sfermion coannihilation processes. 

In this analysis we consider the DM sector of the MSSM, where we take 
into account the bino $M_1$, wino $M_2$, and higgsino $\mu$ mass parameters. 
The $\mu$-term is assumed to be positive. 
Furthermore, the LHC multijet plus missing energy search
\cite{Chatrchyan:2014lfa} can put a very stringent mass limit on the 
gluino mass $M_3$ and the squark masses. To avoid this limit, we take 
the gluino mass to be the same as the universal sfermion masses 
$m_{\tilde{f}}$ that are heavier than $\max[M_1,M_2,\mu,m_A, 800\gev]$.
The pseudo-scalar Higgs mass $m_A$ and the value of $\tan\beta$ control 
the higgsino mixing and the A-resonance region which can have higher 
annihilation cross sections into monochromatic $\gamma$-rays. 
We unify all the trilinear couplings to be $A_0$. Finally, we show the 
ranges of the MSSM7 parameters used in our scan:
\begin{eqnarray}
&&3<\tan\beta<62,\nonumber \\
&&10 <  (M_1,M_2,\mu)/\rm{GeV}  < 4000, \nonumber\\
&& 200<m_A/\rm{GeV}<8000,\nonumber\\
&& \max[M_1,M_2,\mu,m_A,800.0   \gev]<m_{\tilde{f}}<8000\gev,\nonumber\\
&& M_3=m_{\tilde{f}},\nonumber\\
&& -5 < A_0/\rm{TeV} <5.
\end{eqnarray}   
   
% --- relic density
We assume a non-thermal relic scenario in which the neutralino 
can be reproduced by late decays from other particles. The advantage of 
this scenario is to allow for a smaller relic density during the
thermal freeze-out stage, and hence a larger interaction cross section
of DM. The late time production of DM can be responsible to the
measured relic density today. The upper limit of relic density is taken 
from the PLANCK measurement~\cite{Ade:2013zuv}. Besides the relic density, 
we also consider some other common constraints. 
% --- chargino ---
We only allow chargino masses greater than 103.5 GeV~\cite{LEP2}. 
% ---Higgs mass 1503.07589 : 125.09 \pm 0.21 (stat.) \pm 0.11 (syst.)
The Higgs mass is constrained using the most recent combined analysis 
from ATLAS and CMS~\cite{Aad:2015zhl} taking into account 2 GeV 
theoretical uncertainties in the Higgs likelihood function.
% --- invisible higgs decay (1412.8662)
% --- invisible z decay PDG
In the neutralino mass region close to the $Z/h$ mass, we consider the 
$Z$ and Higgs invisible decays~\cite{Agashe:2014kda}.
% --- bs gamma = 3.43e-4\pm0.22e-4\pm 0.21e-4
% --- bs mu mu = 2.9e-9\pm 0.7e-9
Moreover, one can the use the measurement of $\bsm$ and $\bsg$ to further 
constrain the Higgs sector. Here we use the updated data $\brbsg\times 
10^{4}=3.43\pm 0.22\pm 0.21$~\cite{bsg} and $\brbsm\times 10^{9}=2.9\pm 
0.7$~\cite{Aaij:2013aka}. In the $\bsm$ likelihood function we also 
include $10\%$ theoretical uncertainties. 
% --- DD ---
Regarding DM direct detection which can put a stringent limit on the
DM parameter space, we use the spin-independent cross section limit from 
LUX~\cite{Akerib:2013tjd}, the DM-proton spin-dependent cross section 
limit from PICO-2L~\cite{Amole:2015lsj} and the DM-neutron spin-dependent 
cross section limit from XENON100~\cite{Aprile:2013doa}. 

The scan of the parameter space is performed using the package 
\texttt{MultiNest} \cite{Feroz:2008xx}, which is optimized for 
the Bayesian sampling. The scans are driven by the likelihood function 
with all the above constraints. We use 15000 live points for
the sampling. The evidence tolerance factor is $10^{-2}$, and the sampling 
efficiency is 0.8. To obtain a good coverage of the parameter space,
we combine six separate scans, three of which have log priors of the
mass parameters and the other three have flat priors. The SUSY mass 
spectrum is computed by using \texttt{SOFTSUSY}~\cite{Allanach:2001kg} and 
then passed to \texttt{DarkSUSY}~\cite{Gondolo:2004sc} for DM observables, 
\texttt{SuperIso}~\cite{Mahmoudi:2008tp} for $\bsm$ and $\bsg$ computation, 
and \texttt{SUSY-HIT}~\cite{Djouadi:2006bz} for Higgs decay width. 
In addition, we have checked all our collected points under the package 
\texttt{HiggsBounds}~\cite{Bechtle:2013wla} to ensure that we do not 
violate any existing Higgs constraints.

\begin{figure*}[!htb]
\centering
\includegraphics[width=0.48\textwidth]{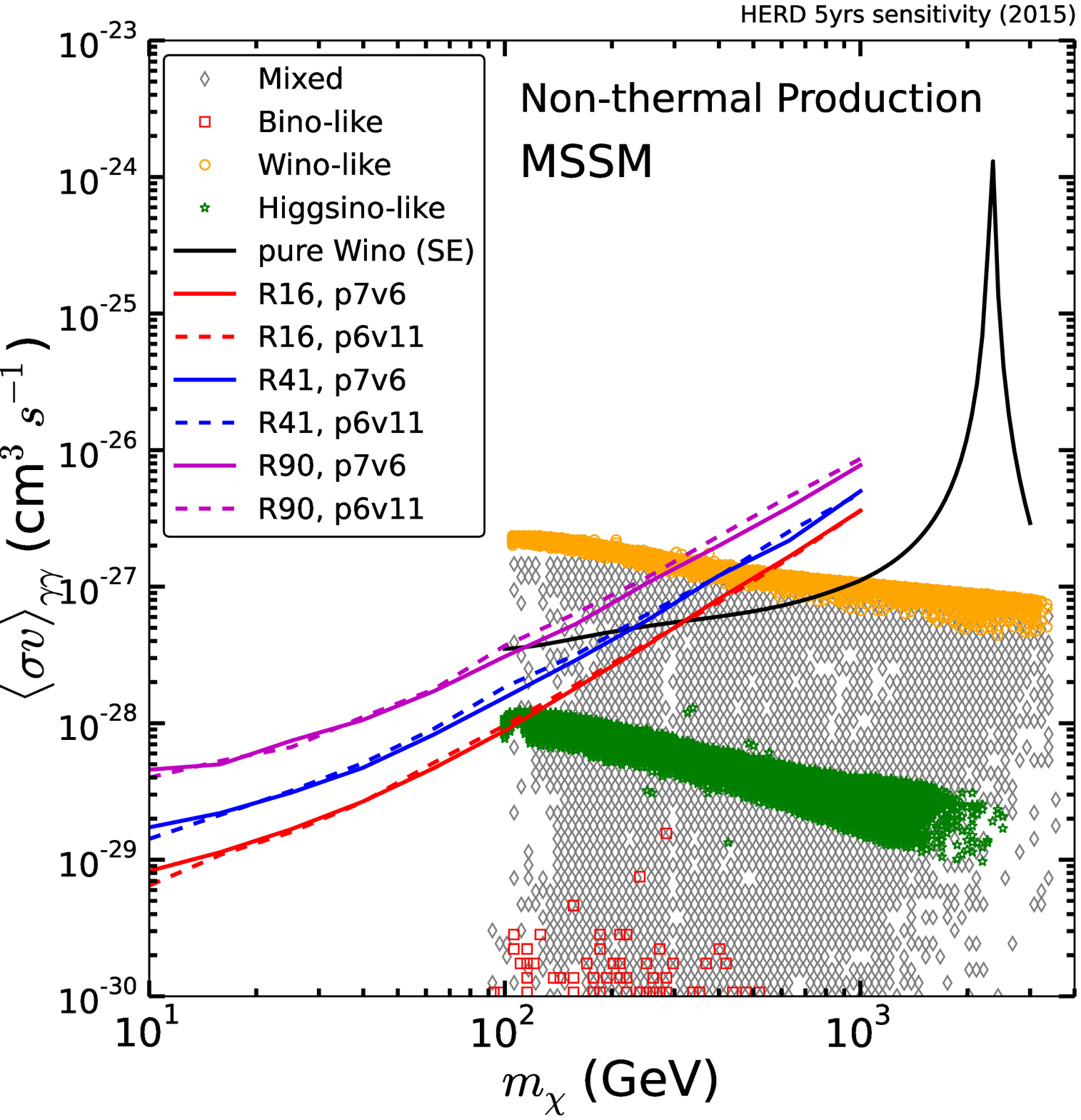}
\includegraphics[width=0.48\textwidth]{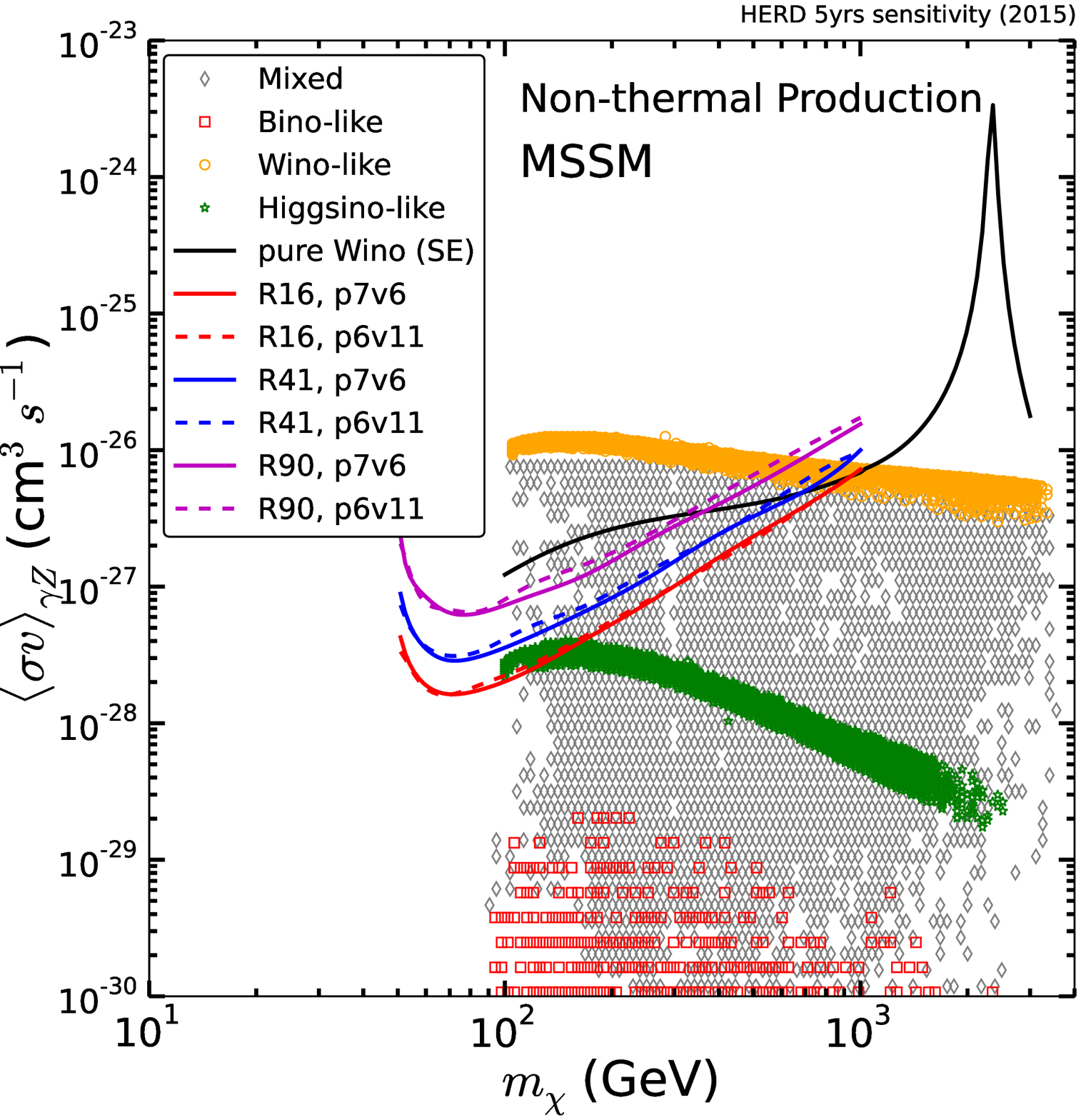}
\caption{
The symbols denote the scattering points in the $m_\chi$-$\langle \sigma v \rangle$ parameter space that satisfy the 
$2\sigma$ ranges (limits) of all the constraints described in the text.
The bino- (red squares), wino- (orange cycles), and higgsino- (green stars) 
like neutralinos are defined by the neutralino composition fraction, 
$g_i>0.95$. In contrast, the mixed neutralino ($g_i<0.95$) is shown in 
gray diamonds. The Sommerfeld enhancement for pure wino ($g_{\rm{wino}}=1$) 
is presented here as the black line for reference. The HERD 5-yr 
sensitivities as shown in Fig.~\ref{fig:results} are over-plotted.
The left panel is for $\gamma\gamma$ channel, and the right panel is
for $\gamma Z$ channel, respectively. We would like to emphasize that in our analysis we assumed a non-thermal relic scenario, where the neutralino can be reproduced by late decays 
from other particles. However, we require that dark matter is not thermally overproduced in the early Universe.
\label{fig:MSSM}}
\end{figure*}

It is easier to identify the neutralino features by considering bino-like, 
wino-like, higgsino-like and mixed neutralinos. The neutralino $\chi^0_1$ 
is decomposed into bino, wino, and higgsinos as
\[
\chi^0_1=Z_{\rm{bino}} \tilde{B} +Z_{\rm{wino}} \tilde{W}+
Z_{H_u} \tilde{H_u} + Z_{H_d} \tilde{H_d}\,.
\]
The fraction of each component, $g_i$  where $i$ denotes bino, wino, or 
higgsino, is defined as $g_{\rm{bino}} =Z^{2}_{\rm{bino}}$, 
$g_{\rm{wino}}=Z^{2}_{\rm{wino}}$, and $g_{\rm{higgsino}}=Z_{H_u}^2 + 
Z_{H_d}^2$. We, therefore, identify the neutralino composition as bino-, 
wino- or higgsino-like when the corresponding fraction $g_i>0.95$.
The other linear combinations are  denoted mixed neutralinos.

In Fig.~\ref{fig:MSSM}, we show the $2\sigma$ allowed scattering points 
for all the constraints described above, where for the relic 
density the $2\sigma$ upper limit is applied. The red, green, and purple 
lines are the HERD sensitivity for the R16, R41, and R90 regions
(see Fig.~\ref{fig:mask_line}). Note that the $Z$-boson  and SM Higgs boson 
resonance  do not appear in both plots since they have very  low 
$\gamma\gamma$ and $\gamma Z$ annihilation cross sections in the 
MSSM~\cite{Hooper:2003ka}. In the neutralino mass region between the 
Higgs resonance region ($m_\chi\simeq m_h/2$) and the chargino-neutralino 
coannihilation region $m_\chi\simeq m_{\chi^\pm}>100.0\gev$, the DM 
relic density is over produced due to the absence of reduction 
mechanisms for the annihilation cross section at freeze out. 
Although we do not include the Sommerfeld enhancement (SE) in our 
cross section computation, we show the SE for pure wino ($g_{\rm{wino}}=1$) 
with the black line as a reference. Note that our wino-like (orange) 
points do not denote wino-wino annihilations only. In this region the 
winos are mixed with some small fraction of higgsinos so that the cross 
section can be higher than the pure wino annihilation cross section 
(black line), in particular in the low mass region. One can see that,
even without the SE, HERD is able to probe the wino region with
$m_\chi<500\gev$ in the $\gamma\gamma$ search and $m_\chi<1\tev$ in the 
$\gamma Z$ search. The wino parameter space can be further constrained 
if SE is included.  The bino region on the other hand is still hard to be probe. 
Note, however, that the points in parameter space where dark matter is entirely produced via thermal freeze-out
are not in the reach of the HERD instrument.

\section{Conclusion}
In this work, we discuss the capability of $\gamma$-ray line searches
of HERD onboard China's space station. Based on the detailed simulations
of the detector performance, including the energy resolution, effective
area and exposure, and the background rejection power, the sensitivity
of monochromatic photon detection of HERD is investigated with the maximum
likelihood analysis. Different DM density profiles, and hence different
optimal ROIs, are discussed. We find that the electrons (dominant at low
energies) and Galactic diffuse $\gamma$-rays (dominant at high energies)
constitute the main contributions to the background. Accordingly the 
results are more uncertain at low energies ($\lesssim100$ GeV) due to 
the different ROIs for the different DM density profiles. Compared with 
the current Fermi results, HERD would improve the limits by a factor of 
a few, for similar observation time. This is mainly due to the largely 
improved energy resolution of HERD ($\sim1\%$). Compared to ground based 
IACTs such as CTA, HERD will play a complementary role in $\gamma$-ray 
line searches. For energies below a few hundred GeV HERD will be more 
sensitive than the IACTs.

\section*{Acknowledgements}
X.H.~and A.L.~would like to thank Alejandro Ibarra for helpful discussions and comments on this article. 
Y.S.T.~would like to thank Shigeki Matsumoto and Satyanarayan Mukhopadhyay for helpful discussions. Y.S.T.~was supported by World Premier International Research Center 
Initiative (WPI), MEXT, Japan. This research was partially supported by 
the by the Graduiertenkolleg "Particle Physics at the Energy Frontier 
of New Phenomena" and by the TUM Graduate School.

\bibliographystyle{apsrev}
\bibliography{refs}

\end{document}